\documentclass[12pt]{iopart}

\usepackage{iopams}
\usepackage{graphicx}

\begin{document}

\title[]{Electromechanically induced absorption in a circuit nano-electromechanical system}

\author{Fredrik Hocke$^1$, Xiaoqing Zhou$^{2,4}$, Albert Schliesser$^{2,4}$, Tobias J. Kippenberg$^{2,4}$ Hans Huebl$^1$, and Rudolf Gross$^{1,3}$}

\address{$^1$ Walther-Mei{\ss}ner-Institut, Bayerische Akademie der Wissenschaften, D-85748 Garching, Germany}
\address{$^2$ \'{E}cole Polytechnique F\'{e}d\'{e}rale de Lausanne (EPFL), CH-1015 Lausanne, Switzerland}
\address{$^3$ Physik-Department, Technische Universit\"{a}t M\"{u}nchen, D-85748 Garching, Germany}
\address{$^4$ Max-Planck-Institut f\"{u}r Quantenoptik, D-85748 Garching, Germany}
\eads{\mailto{fredrik.hocke@wmi.badw.de}, \mailto{rudolf.gross@wmi.badw.de}}

\begin{abstract}
  A detailed analysis of electromechanically induced absorption (EMIA)  in a circuit nano-electromechanical hybrid system consisting of a superconducting microwave resonator coupled to a nanomechanical beam is presented. By performing two-tone spectroscopy experiments we have studied EMIA as a function of the drive power over a wide range of drive and probe tone detunings. We find good quantitative agreement between experiment and theoretical modeling based on the Hamiltonian formulation of a generic electromechanical system. We show that the absorption of microwave signals in an extremely narrow frequency band ($\Delta\omega/2\pi <5$\,Hz) around the cavity resonance of about 6\,GHz can be adjusted over a range of more than $25$\,dB on varying the drive tone power by a factor of two. Possible applications of this phenomenon include notch filters to cut out extremely narrow frequency bands ($<$\,Hz) of a much broader band of the order of MHz defined by the resonance width of the microwave cavity. The amount of absorption as well as the filtered frequency is tunable over the full width of the microwave resonance by adjusting the power and frequency of the drive field. At high drive power we observe parametric microwave amplification with the nanomechanical resonator. Due to the very low loss rate of the nanomechanical beam the drive power range for parametric amplification is narrow, since the beam rapidly starts to perform self-oscillations.
\end{abstract}

\pacs{85.85.+j, 42.50.Gy, 42.50.Wk, 84.40.Dc}

\submitto{New J. Phys.}
\maketitle

\section{Introduction}
\label{Introduction}

The combination of cavities for electromagnetic radiation with engineered mechanical resonators to form cavity electromechanical systems allows for quantum control over mechanical motion or, conversely, mechanical control over electromagnetic fields. Depending on the frequency of the electromagnetic field, this class of hybrid systems forms the new fields of cavity opto- or electromechanics \cite{Kippenberg2007, Kippenberg2008, Florian2009, Aspelmeyer2010, Regal2011, Aspelmeyer2012}. Cavity opto- and electromechanical systems allow us to study a wealth of interesting physical phenomena and to design novel devices that can probe extremely tiny forces \cite{Chaste2011}, which may enable the mechanical detection and imaging of a single electron spin, or allow controlling the quantum state of mechanical oscillators. In these systems consisting of a mechanical oscillator affected by the radiation pressure force induced by the electromagnetic field of an optical or microwave cavity, parametric electromechanical coupling occurs. This parametric coupling makes both amplification and cooling \cite{Schliesser2006, Gigan2006, Linthorne1990} of the mechanical mode possible. The latter allowed several groups to approach the quantum mechanical ground state of the mechanical system \cite{Groblacher2009, Park2009a, Schliesser2009, Rocheleau2010, Riviere2011} and even achieve mechanical oscillators with high ground state probability \cite{O'Connell2010, Teufel2011a, Chan2011, Verhagen2012} and active quantum coherent coupling \cite{O'Connell2010, Teufel2011a, Verhagen2012}. Further important achievements in the field of cavity opto- and electromechanics are, among others, the access to the regime of strong coupling \cite{Groblacher2009a, Teufel2011},
and the measurement of nanomechanical motion with an imprecision below that of the standard quantum limit \cite{Teufel2009, Anetsberger2009, Anetsberger2010}. An important ingredient for these experiments is the application of two-tone spectroscopy. Here, the system is driven by a strong drive tone at angular frequency $\omega_{\rm d}$, which is red- or blue-detuned from the cavity resonant angular frequency $\omega_{\rm c}$ by about the mechanical resonant angular frequency $\Omega_{\rm m}$, while a weak probe tone at angular frequency $\omega_{\rm p}$ probes the modified cavity resonance. Two-tone spectroscopy has been successfully used for the study of opto-/electromechanically induced transparency (OMIT/EMIT) or absorption (OMIA/EMIA) \cite{Weis2010,Safavi-Naeini2011}. Here, the electromagnetic response of the system to a weak probe field is controlled by the drive field, driving the lower or upper motional sideband, respectively. The resulting destructive or constructive interference of different excitation pathways of the intracavity field results in a narrow window of enhanced or reduced transparency when the two-photon resonance condition is met. Both EMIT and EMIA are key requirements for protocols to manipulate and control electromagnetical signals \cite{Weis2010, Teufel2011, Safavi-Naeini2011, Massel2011}. These, in turn, are prerequisites for the realization of tools e.g. for the conditioning of microwave signals working on the quantum level. To this end, experiments on the red-detuned sideband can become limited by the linewidth of the mechanical resonator, which is broadened by the opto- or electromechanical damping. On the other hand, experiments on the blue-detuned sideband are not restricted by this effect. Here, the filter function can become even more narrow due to electromechanical linewidth narrowing. Hence, in combination with the demonstration of EMIA \cite{Safavi-Naeini2011} and parametric amplification \cite{Massel2011}, applications are within reach.

In this article, we present a detailed study of EMIA in a hybrid system consisting of a Nb superconducting microwave resonator coupled to a high Q nanomechanical beam. We show that the absorption of microwave signals at cavity resonance can be increased by more than $25$\,dB on increasing the power of the drive tone by a factor of two. In order to obtain detailed understanding of the absorption effect as a function of the drive and the probe frequency, we performed systematic two-tone spectroscopy experiments. We find the appearance of an absorption double dip as a function of the drive frequency. We explain these findings quantitatively within the model of electromechanically induced absorption. Furthermore, in the limit of high drive power we observe that the absorbtion depth decreases again and even reverses sign, therefore showing signatures similar to those known from EMIT. At the largest drive powers, parametric amplification and self-oscillation \cite{Kippenberg2005, Marquardt2006} is observed.

\section{Electromechanical induced absorption}
\label{Electromechanical induced absorption}

Cavity electromechanical systems incorporating a parametrical coupling which links the position of a mechanical oscillator to the photon number of an electromagnetic resonator can be described by the Hamiltonian \cite{Wilson-Rae2007, Marquardt2007a}
\begin{equation}
\hat{H} = \hbar\widetilde{\omega}_{\rm c}\left(\hat{n}_{\rm c}+\frac{1}{2}\right) + \hbar\Omega_{\rm m} \left(\hat{n}_{\rm m}+\frac{1}{2}\right) + \hbar G \hat{n}_{\rm c} \hat{x} + \hat{H}_{\rm d} \; .
\label{equ:Hamiltion}
\end{equation}
Here, $\hat{n}_{\rm c} = \hat{a}^\dag\hat{a}$ and $\hat{n}_{\rm m} = \hat{b}^\dag\hat{b}$ are the intra-cavity (photon) and mechanical mode (phonon) excitation number operators with $\hat{a}^\dag$ ($\hat{a}$) and $\hat{b}^\dag$ ($\hat{b}$) the photon and phonon creation (annihilation) operators, respectively, $\hbar$ is the reduced Planck constant, $\hat{x}=x_{\rm{zp}}(\hat{b}^\dag+\hat{b})$ the mechanical displacement operator with $x_{\rm{zp}}=\sqrt{\hbar/2m_{\rm{eff}}\Omega_{\rm{m}}}$ the cantilever zero-point fluctuation and $m_{\rm{eff}}$ the mechanical oscillator mass, $g_0 = Gx_{\rm{zp}}$ the vacuum electromechanical coupling rate \cite{Gorodetsky2010a}, and $\hat{H}_{\rm{d}}$ represents the external driving fields. The quantity $G = d\widetilde{\omega}_{\rm{c}}/dx$ characterizes the linear electromechanical interaction between the mechanical mode and the cavity field, as the cavity resonance frequency varies as $\omega_{\rm c} = \widetilde{\omega}_{\rm c} + G\bar{x}$. The linear shift $G \bar{x}$ of the bare cavity eigenfrequency $\widetilde{\omega}_{\rm{c}}$ is caused by the average static displacement $\bar{x}=\langle \hat{x} \rangle$ of the mechanical resonator due to the static radiation pressure of the electromagnetic field \cite{Dorsel1983}. We note that in a detailed discussion the renormalized cavity resonance frequency $\omega_{\rm{c}}=\widetilde{\omega}_{\rm{c}}+G \bar{x}$ has to be used. However, in most cases $G \bar{x}$ is small and one can use $\omega_{\rm{c}} \simeq \widetilde{\omega}_{\rm{c}}$. For our system, where a mechanical beam is capacitively coupled to a microwave coplanar waveguide resonator [cf. figure~\ref{Hocke_EMIA_Figure2}(a)], the electromechanical interaction is given by the change of the overall capacitance of the resonator due to the displacement of the mechanical beam.

In the rotating wave approximation the interaction Hamiltonian $\hat{H}_{\rm int} = \hbar G \hat{n}_{\rm c} \hat{x} = \hbar g_0 \hat{a}^\dag\hat{a} (\hat{b}^\dag+\hat{b})$ can be simplified. Under red-detuned driving it becomes a beam splitter like Hamiltonian $\hat{H}_{\rm int} = \hbar g (\hat{a}\hat{b}^\dag + \hat{a}^\dag\hat{b})$, whereas under blue-detuned driving it can be written as $\hat{H}_{\rm int} = \hbar g (\hat{a}^\dag\hat{b}^\dag + \hat{a}\hat{b})$, corresponding to the one of parametric amplification. Here, $g = g_0 \sqrt{\bar{n}_{\rm c}}$ is the field-enhanced coupling rate \cite{Wilson-Rae2007, Marquardt2007a}, with $\bar{n}_{\rm c}$ the average number of photons in the cavity. Until today, a large variety of physical implementations described by the above Hamiltonian have been realized. While cavities both in the optical and microwave domain are used, the mechanical systems range from gram scale mirrors \cite{Corbitt2007a} to nanomechanical beams with an effective mass of picograms \cite{Regal2008}.

\begin{figure}[tb]
\center{\includegraphics[width=0.8\columnwidth]{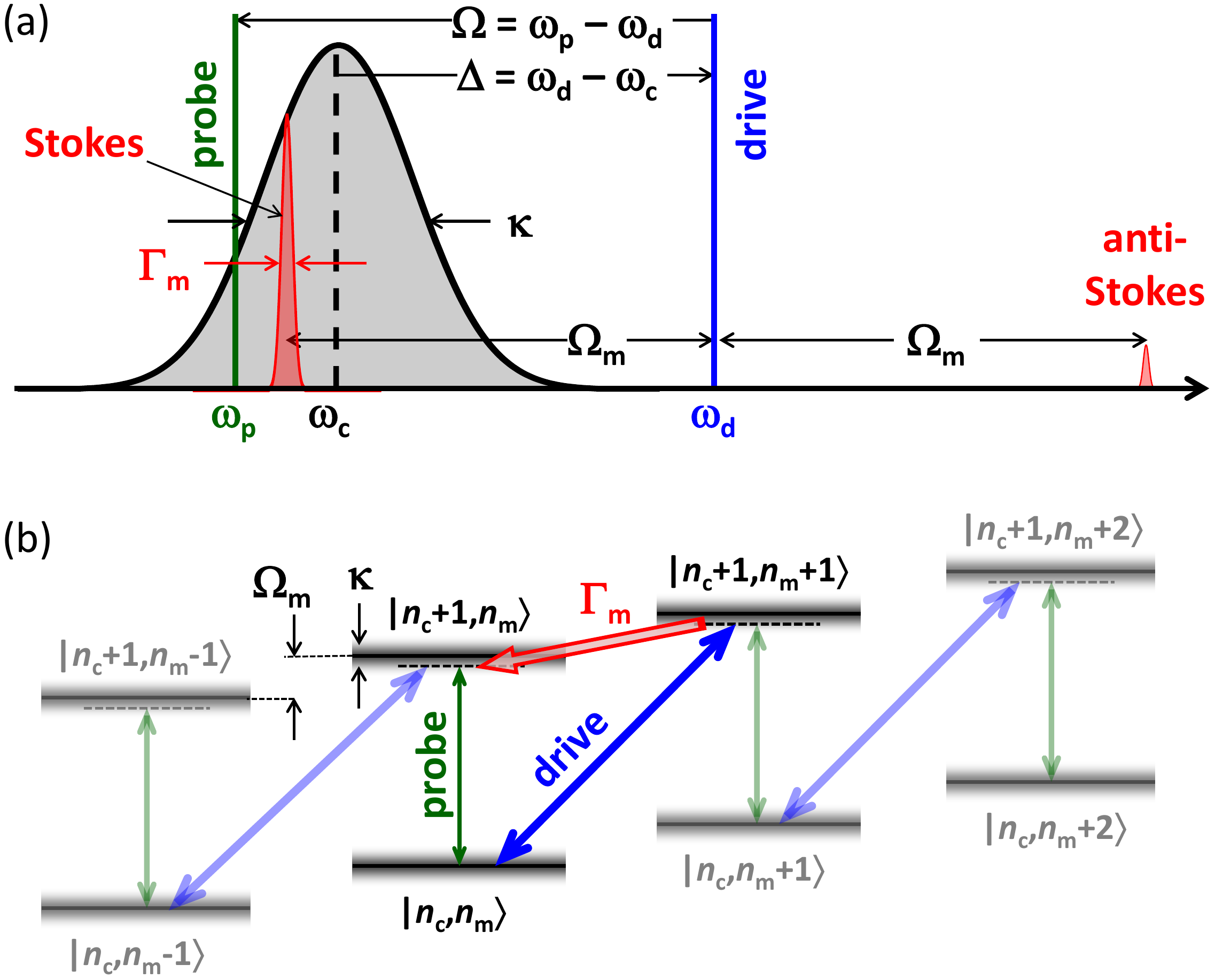}}
 \caption{
  (a) Schematic diagram explaining the various frequencies:  driving tone frequency $\omega_{\rm d}$ (blue line), probe tone frequency $\omega_{\rm p}$ (green line), cavity resonant frequency $\omega_{\rm c}$ (black line), drive tone detuning $\Delta = \omega_{\rm d}-\omega_{\rm c}$, probe tone detuning $\Omega = \omega_{\rm p}-\omega_{\rm d}$. Varying $\Delta$ shifts the drive tone with respect to the cavity resonance (grey-shaded area). Since the Stokes line (red shaded area) is fixed to the drive tone at $\omega_{\rm d} - \Omega_{\rm m}$, the variation of $\Delta$ also shifts the Stokes line with respect to the cavity resonance. Varying $\Omega$ at constant $\Delta$ shifts the probe tone with respect to the cavity resonance and the Stokes line. (b) Level diagram describing the physics of the electromechanical system: The blue detuned drive tone at frequency $\omega_{\rm d}$ (blue arrow) induces a transition from a state $|n_{\rm c}, n_{\rm m}\rangle$ characterized by $n_{\rm c}$ cavity excitations (photons) and $n_{\rm m}$ mechanical excitations (phonons) to a state $|n_{\rm c}+1, n_{\rm m}+1\rangle$, that is, it adds one excitation both to the cavity and the mechanical oscillator. The cavity photons generated in this down-conversion process are degenerate with the near-resonant probe field at frequency $\omega_{\rm p}=\omega_{\rm d}-\Omega_{\rm m}$ (green arrow), resulting in a constructive interference effect. The dashed lines indicate virtual levels, indicating that the drive field is not optimally blue-detuned from the cavity resonance ($\Omega = \Omega_{\rm m} \neq \Delta$).
 }
\label{Hocke_EMIA_Figure1}
\end{figure}

Driving the electromechanical system at angular frequency $\omega_{\rm d}$, the sign of the detuning $\Delta=\omega_{\rm{d}}-\omega_{\rm{c}}$ of the drive tone determines the direction of energy transfer between the electromagnetic and mechanical system. A red- and blue-detuned drive tone, i.e. $\Delta <0$ and $\Delta > 0$, allows for parametric cooling and amplification of the mechanical system, respectively. In the resolved-sideband regime \cite{Schliesser2008}, where the mechanical frequency exceeds the cavity loss rate, $\Omega_{\rm m} > \kappa$, parametric cooling (amplification) is most effective for $\Delta =\sqrt{\Omega_{\rm{m}}^2+(\kappa /2)^2}$, that is, if the drive tone is red (blue) detuned close to the lower (upper) motional sideband, $\Delta=\omega_{\rm{d}}-\omega_{\rm{c}} \approx -\Omega_{\rm{m}}$ ($+\Omega_{\rm{m}}$) \cite{Wilson-Rae2007,Marquardt2007a}. We will refer to this case as optimum detuning.

In the following, we consider only the case of blue-detuning ($\Delta \simeq + \Omega_{\rm{m}}$). In this case a strong drive tone at $\omega_{\rm d}$ close to the upper motional sideband of the cavity is applied to the system, while a second, much weaker tone probes the modified cavity resonance at frequency $\omega_{\rm{p}}=\omega_{\rm{d}}+\Omega$ [cf. figure~\ref{Hocke_EMIA_Figure1}(a)]. The simultaneous presence of the drive and the probe tone result in a radiation pressure force oscillating at $\Omega = \omega_{\rm{p}}-\omega_{\rm{d}}$. If this difference frequency is close to the mechanical resonance frequency, $\Omega \simeq -\Omega_{\rm m}$, a coherent oscillation of the mechanical system is induced. As a consequence of this oscillation, Stokes and anti-Stokes fields [red-shaded areas in figure~\ref{Hocke_EMIA_Figure1}(a)] build up at $\omega_{\rm d} \pm \Omega_{\rm m}$ around the strong driving field. The microwave resonator acts as a narrow-band filter for these fields. If the system is in the resolved-sideband regime, $\Omega_{\rm m} > \kappa$, the anti-Stokes line at $\omega_{\rm d} + \Omega_{\rm m} > \omega_{\rm c}$ is strongly suppressed because it is off-resonant with the cavity, whereas the Stokes line at $\omega_{\rm d} - \Omega_{\rm m} \simeq \omega_{\rm c}$ is enhanced. Moreover, since the Stokes scattered field is degenerate with the probe field sent to the cavity, it allows for constructive interference of the two fields enhancing the build-up of the intra-cavity probe field. This can be viewed as a self-interference between two different excitation pathways. The resulting increased feeding (``absorption'') of probe photons into the cavity 
manifests itself as a reduced cavity transmission. Depending on the frequency of the electromagnetic field, this effect is referred to as electromechanically or optomechanically induced absorption (EMIA or OMIA) \cite{Safavi-Naeini2011}, the electromechanical analog of electromagnetically induced absorption (EIA) \cite{Lezama1999}. At even higher drive field power, the system switches from EMIA to parametric amplification \cite{Massel2011}, resulting in electromagnetic signal amplification, and eventually phonon-lasing. In the picture of parametric amplification, the electromechanical system can be viewed as a parametric amplifier strongly pumped at $\omega_{\rm d}$ and amplifying the weak input signal at $\omega_{\rm p}$. This amplification is detected as an enhanced cavity transmission.

Another way of visualizing the effect of EMIA is the use of the level scheme shown in figure~\ref{Hocke_EMIA_Figure1}(b). In this scheme the electromechanical states consist of product states $|n_{\rm c}, n_{\rm m}\rangle$ characterized by $n_{\rm c}$ cavity excitations (photons) and $n_{\rm m}$ mechanical excitations (phonons). Here, a pure cavity excitation would be represented by a vertical arrow increasing only the number photons in the cavity. In contrast, if the drive field is set to $\omega_{\rm d} \simeq \omega_{\rm c}+\Omega_{\rm m}$ (optimal detuning), it increases both the photon and the phonon number by one and hence is represented by a diagonal arrow. The photons of the drive field are down-converted and scatter into the Stokes line at $\omega_{\rm d}- \Omega_{\rm m}$, matching approximately the cavity resonant frequency $\omega_{\rm c}$. Since probe and drive tone are coherent with respect to each other\footnote{Note that for EMIA the mutual coherence of probe and drive needs to be reciprocal of the absorption window.}, they can interfere with the weak probe field, indicated by the vertical green arrow. In the case that the drive field is not optimally detuned from the cavity resonance ($\Delta = \omega_{\rm d} - \omega_{\rm c} \neq \Omega_{\rm m}$), virtual levels are involved which are represented by the dashed lines in Fig.~\ref{Hocke_EMIA_Figure1}(b). Also in this situation the down-conversion of the drive field photons generates photons of frequency $\omega_{\rm d} -\Omega_{\rm m} = \omega_{\rm p} \neq \omega_{\rm c}$, that is, photons of the same frequency as the probe field, again allowing for constructive interference with the probe field sent to the cavity. Depending on the relative power of the probe and the drive field, the interference can lead to a partial or full extinction of the probe field outside the cavity, what is detected as a reduced cavity transmission (enhanced absorption) in experiments.

For a quantitative analysis of the absorption as a function of the drive power and the detunings $\Delta$ and $\Omega$, one has to solve the full Hamiltonian (\ref{equ:Hamiltion}), including both the drive and probe fields as well as the losses in the electromagnetic and mechanical resonator. Fortunately, the weak probe field allows for a linearization of the system's dynamics around the steady-state values $\bar{n}_{\rm c} = \langle \hat{n}_{\rm c}\rangle$ and $\bar{x}=\langle \hat{x}\rangle$, where the former is the average number of photons inside the cavity and the latter the average displacement of the mechanical mode. Solving the resulting Langevin equations for the intra-cavity field in a frame rotating with $\Delta$ and calculating the transmission spectrum of the probe field using input-output theory \cite{Weis2010, Safavi-Naeini2011, Agarwal2010, Gardiner2004}, yields the following expression for the probe power transmission
\begin{equation}
|t|^2 = S_{\rm{out}} = \left| 1-\frac{\kappa_{\rm{ex}}/2}{-i(\Delta +\Omega )+\kappa/2+\frac{g_{0}^2 \bar{n}_{\rm c}}{i(\Omega+\Omega_{\rm{m}})-\Gamma_{\rm{m}}/2}} \right|^2 \; .
\label{equ:S21_approx}
\end{equation}
Here, $g = g_{0} \sqrt{\bar{n}_{\rm c}}$ is the field-enhanced electromechanical coupling rate, $\Gamma_{\rm m}$ the loss rate of the mechanical oscillator, and $\kappa/2\pi = (\kappa_{\rm{ex}} + \kappa_{\rm{in}})/2\pi$ the total loss rate of the cavity taking into account both internal losses due to dissipative effects and external losses due to the finite coupling to the feedline. In deriving eq.(\ref{equ:S21_approx}), the strong suppression of the anti-Stokes field due to the filter function of the electromagnetic resonator has been taken into account. We also note that the transmission spectrum is completely analogous to that obtained for EIA. However, in contrast to atomic systems the coupling strength between the electromagnetic and mechanical mode can be easily varied over a wide range by changing the average photon number $\bar{n}_{\rm c}$ inside the cavity.

\section{Experimental techniques}

\begin{figure}[tb]
\center{\includegraphics[width=0.8\columnwidth]{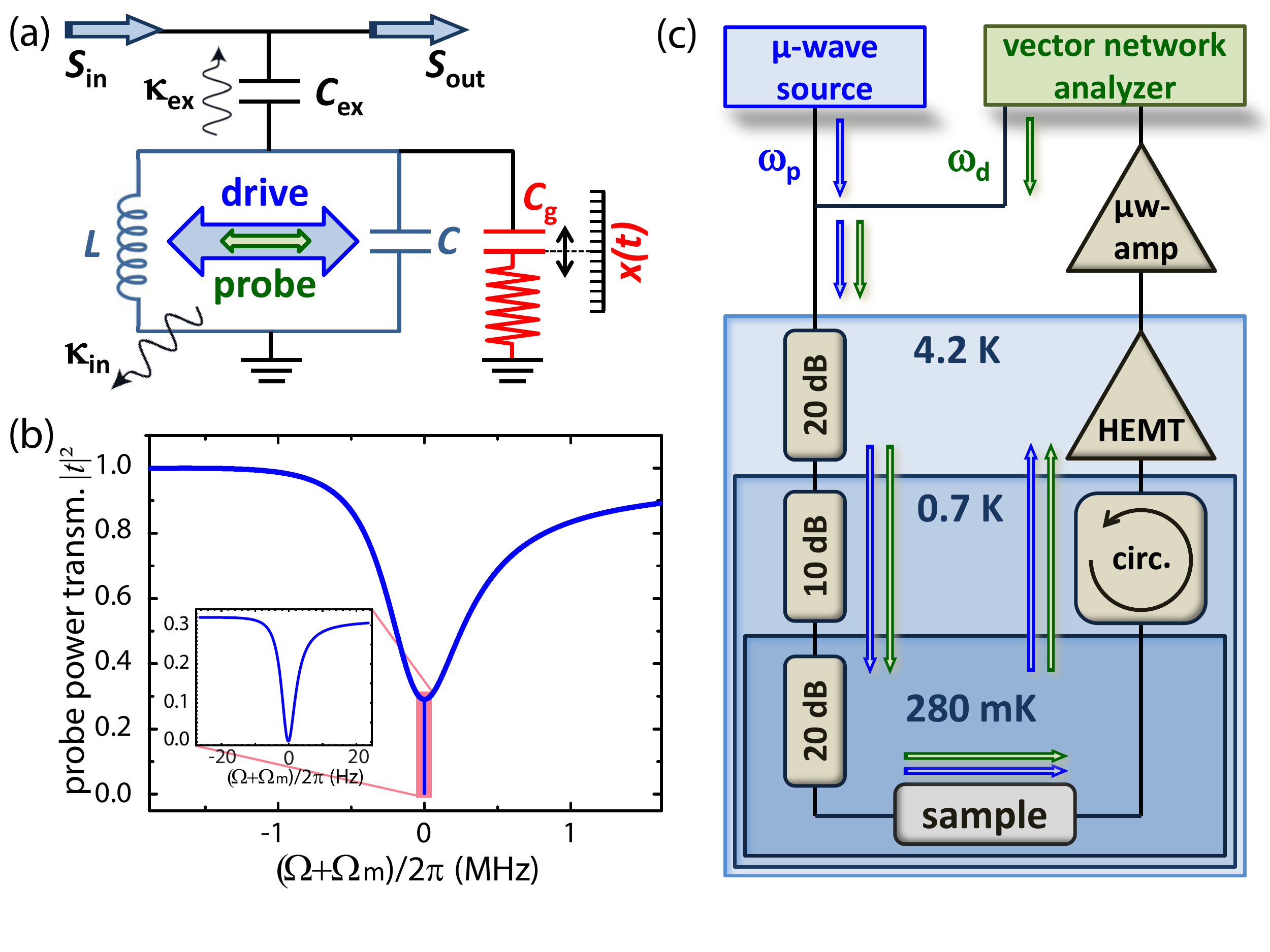}}
\caption{
  (a) Equivalent circuit of the nano-electromechanical device under investigation. The variable capacitance $C_{\rm g}$ due to the oscillating nanobeam is parallel to the capacitance $C$ of the cavity represented by a lumped element $LC$ resonator with internal loss rate $\kappa_{\rm in}$. The cavity is coupled to the feedline via the capacitance $C_{\rm ex}$, resulting in the external loss rate $\kappa_{\rm ex}$. (b) Schematic of a typical probe power transmission spectrum of a superconducting CPW microwave resonator with an optimally detuned driving tone applied at $\Delta = \Omega_{\rm m}$. The inset shows the very sharp absorption feature close to $\Omega+\Omega_{\rm m}=0$ on an enlarge scale. (c) Scheme of the experimental setup. A drive tone at $\omega_{\rm d}$ is applied using a microwave source, a probe tone at $\omega_{\rm p}$ is added using a power divider. Both tones are fed into the dilution fridge and directed to the sample via heavily attenuated coaxial cables. The transmitted probe tone is amplified by a cold HEMT and a room-temperature microwave amplifier and analyzed by a vector network analyzer. A circulator is used to isolate the sample from the HEMT amplifier.
  }
 \label{Hocke_EMIA_Figure2}
\end{figure}

The circuit nano-electromechanical device studied in our experiments is a hybrid system, consisting of a superconducting coplanar waveguide (CPW) microwave resonator capacitively coupled to a nanomechanical beam. Since the oscillation of the beam results in an oscillation of the total capacitance of the CPW resonator and hence its resonant frequency, a parametric coupling of the microwave resonator to the vibrational mode of a nanomechanical oscillator is realized. Our nano-electromechanical system is similar to that studied by Regal {\emph{et al.} \cite{Regal2008}. However, in contrast to the purely metallic nanobeams (Al) in \cite{Regal2008}, we use nanobeams consisting of Nb/$\rm{Si}_{3}\rm{N}_{4}$ bilayers to increase the resonant frequency and the quality factor of the mechanical oscillator. Furthermore, the superconducting CPW resonator is made of Nb with a high critical temperature of $9.2$\,K, resulting in a higher internal quality an lager critical photon number of the microwave resonator. The latter is fabricated by patterning a quarter-wave CPW structure into a $200$\,nm thick Nb film, which has been deposited on a silicon substrate by sputter deposition \cite{Day2003, Niemczyk2010, Niemczyk2009}. The resonator is capacitively coupled to a $50\,\Omega$ CPW feedline at one end and shortened at the other (one-sided cavity). By choosing the resonator length an eigenfrequency of $\omega_{\rm{c}}/2\pi = 6.07$\,GHz is obtained. We note that the resonant frequency $\omega_{\rm{c}}$ also depends on the additional capacitances $C_{\rm g}$ and $C_{\rm ex}$ due to the capacitive coupling to the nanomechanical beam and the feedline, respectively [cf. figure~\ref{Hocke_EMIA_Figure2}(a)]. The coupling to the feedline results in a coupling rate of $\kappa_{\rm{ex}}=2\pi\times339$\,kHz. Comparing this coupling rate to the measured cavity linewidth (total loss rate) of $\kappa=2\pi\times759$\,kHz yields a coupling rate $\eta_{\rm c} = \kappa_{\rm{ex}}/\kappa \simeq 0.45$ close to the critical coupling $\eta_{\rm c} = 1/2$ where the best contrast is obtained. A schematic transmission spectrum is shown in Fig.~\ref{Hocke_EMIA_Figure2}(b).

The nanomechanical resonator (NR) consists of a Nb/$\rm{Si}_{3}\rm{N}_{4}$ nanobeam which is clamped on both ends. The Nb/$\rm{Si}_{3}\rm{N}_{4}$ bilayer is made out of a $130$\,nm thick Nb film deposited on top of a $70$\,nm thick, highly strained (tensile) $\rm{Si}_{3}\rm{N}_{4}$ layer. The beam has a high aspect ratio with a length of $60$\,$\mu$m and a width of $140$\,nm. The structure is patterned by electron beam lithography, followed by an anisotropic and an isotropic reactive ion etching process \cite{Zhou2012}. The high tensile strain in the $\rm{Si}_{3}\rm{N}_{4}$ layer accounts for a high mechanical eigenfrequency of $\Omega_{\rm{m}}/2\pi = 1.45$\,MHz and a narrow linewidth of $\Gamma_{\rm{m}}/2\pi = 11$\,Hz \cite{Verbridge2006, Verbridge2008} at $200$\,mK. Unfortunately, the compressive strain in the Nb film partly compensates the tensile strain in $\rm{Si}_{3}\rm{N}_{4}$. This reduces the mechanical eigenfrequency below the value of $5-6$\,MHz expected for pure $\rm{Si}_{3}\rm{N}_{4}$, depending on the detailed amount of the tensile strain \cite{Verbridge2006}. Nevertheless, since $\Omega_{\rm{m}}/2\pi$ is about twice the cavity linewidth $\kappa/2\pi$ ($\Omega_{\rm{m}}/\kappa \simeq 1.91$), the system is sufficiently far in the resolved-sideband regime to be able to neglect the anti-Stokes field in the theoretical modeling. The nanobeam is capacitively coupled to the center conductor of the CPW microwave resonator at the voltage antinode. The finite coupling capacitance $C_{\rm g}$ gives rise to the electromechanical coupling $g_{0} = G x_{\rm{zp}}$ between the mechanical displacement and the microwave mode inside the CPW resonator [cf. figure~\ref{Hocke_EMIA_Figure2}(a)]. For a coupling distance of $200$\,nm, a normalized vacuum coupling $g_{0}=2\pi\times1.26$\,Hz is determined from frequency noise calibration \cite{Gorodetsky2010a, Zhou2012}.
With the zero point fluctuation amplitude of the beam, $x_{\rm{zp}}=\sqrt{\hbar/2m_{\rm{eff}}\Omega_{\rm{m}}} \simeq 30$\,fm ($m_{\rm{eff}}\simeq7$\,pg), the equivalent linear electromechanical interaction can be determined to $G=g_0/x_{\rm zp}=2\pi\times36.3$\,kHz/nm.

As depicted in figure~\ref{Hocke_EMIA_Figure2}(c), the experiments are performed in a dilution refrigerator at a temperature of $T=280$\,mK, far below the critical temperature of Nb ($T_{\rm c} = 9.2$\,K). The microwave excitation and detection circuitry consists of a Rohde\&Schwarz SMF microwave source, used to generate the strong blue-detuned ($\Delta>0$) drive tone at $\omega_{\rm{d}}=\omega_{\rm{c}}+\Delta$, while the weak probe field centered around $\omega_{\rm{d}}-\Omega_{\rm{m}}$ stems from a Rohde\&Schwarz ZVA network analyzer. The latter allows for phase sensitive detection of the the microwave signal after interacting with the device under investigation. For the two-tone experiments, the two microwave signals are combined at room temperature and then sent to the electromechanical hybrid via coaxial cables. Several attenuators at the various temperature stages are used for the reduction of thermal noise and thermalization of the microwave cables. After passing the device under test, the transmitted signal is amplified using a cryogenic low-noise HEMT amplifier anchored at $4.2$\,K and a room-temperature microwave amplifier. The HEMT amplifier is isolated from the sample output by a circulator anchored at $0.7$\,K. To compensate the effect of the attenuators and amplifiers, we normalize the transmitted signal with respect to the transmitted signal away from the cavity resonance.

\section{Experimental results and discussion}

In this subsection we discuss the experimental results of the two-tone spectroscopy experiments performed of our electromechanical system. The inset of figure~\ref{Hocke_EMIA_Figure3} shows a typical probe power transmission spectrum obtained by plotting $|t|^2$ versus the probe tone detuning $\Omega + \Omega_{\rm m}$ for a constant drive power of $P_{\rm d}=264$\,pW applied at $\omega_{\rm{d}}=\omega_{\rm{c}}+\Omega_{\rm{m}}$, that is at optimum drive tone detuning $\Delta = \Omega_{\rm m}$. The spectrum has been recorded for a probe tone power of $2.4$\,fW ($-116.5$\,dBm). At $\Omega=-\Omega_{\rm m}$, a narrow absorption dip is observed in addition to the bare cavity absorption. Note that the much broader absorption curve of the cavity [cf. Fig.~\ref{Hocke_EMIA_Figure2}(b)] is not seen due to the small frequency window of the spectrum shown in the inset of in figure~\ref{Hocke_EMIA_Figure3}. Evidently, the very narrow absorption dip has a reduced mechanical linewidth of $\Gamma_{\rm{eff}}/2\pi = 5.1$\,Hz, which is significantly less than the intrinsic linewidth $\Gamma_{\rm{m}}/2\pi = 11$\,Hz of the mechanical oscillator. The reason is that due to electromechanical coupling the dressed mechanical mode, which is now effectively a phonon-photon polariton, has acquired a finite photonic nature, thereby coupling the mechanical mode to the photonic loss channels at a rate $\gamma = C \Gamma_{\rm m}$. Here, we have introduced the electromechanical cooperativity defined as $C = 4g^2/\kappa\Gamma_{\rm{m}} = 4g_{0}^2\bar{n}_{\rm c}/\kappa\Gamma_{\rm{m}}$ for an optical cavity decay rate of $\kappa$ and an intrinsic mechanical resonance damping rate of $\Gamma_{\rm m}$ \cite{Groblacher2009a}. The additional photonic loss channel causes an electromechanical linewidth narrowing which scales linearly with the average number of drive photons
\begin{equation}
\bar{n}_{\rm c} = \frac{P_{\rm{d}}}{\hbar\omega_{\rm{d}}} \;  \frac{\kappa_{\rm{ex}}/2}{(\kappa/2)^2+\Delta^2} \;
 \label{equ:photonnum}
\end{equation}
and hence with the drive power $P_{\rm{d}}$. Since part of the losses of the mechanical system now occur via the photonic channel, a reduced mechanical linewidth $\Gamma_{\rm{eff}} = \Gamma_{\rm{m}}(1-C)$ is obtained \cite{Weis2010, Teufel2008b}. The cooperativity can easily be accessed in experiments, since it is directly proportional to the drive power. In Fig.~\ref{Hocke_EMIA_Figure3}, the drive power and cooperativity are shown on the bottom and top scale, respectively.

In figure~\ref{Hocke_EMIA_Figure3}, the peak probe power transmission $\left|t_{\rm{0}}\right|^2$ is plotted versus the drive power. It is given by the minimum of the narrow absorption dip shown in the inset of figure~\ref{Hocke_EMIA_Figure3}, which is determined by fitting the data by a damped harmonic oscillator spectrum. The peak probe power transmission decreases with increasing drive power from about 0.3 for the bare cavity minimum ($P_{\rm{d}}=0$) to a minimum value of $\left|t_{\rm{0}}\right|^2=0.0046$ for a drive power of $P_{\rm{d}}=270$\,pW. This corresponds to an additional absorption of $18.3$\,dB with respect to the bare microwave cavity background and $23.4$\,dB with respect to unity transmission. Increasing the drive power further, $\left|t_{\rm{0}}\right|^2$ increases again and exceeds the initial value of the bare cavity, $\left|t_{\rm{0}}\right|^2 (P_{\rm d} =0) \simeq 0.3$, at higher drive powers. In other words, the transmission reduction (``absorption'') turns into a transmission enhancement (``emission''). At drive powers above $380$\,fW, $\left|t_{\rm{0}}\right|^2 <1$, i.e. it even exceeds the undisturbed transmission of the feedline. At this power level we are entering the regime of parametric amplification, which has been studied recently by Massel \emph{et al.} \cite{Massel2011}. Finally, when the drive power is increased even further, the beam starts to perform self-oscillations at $P_{\rm{d}} \simeq 390$\,pW. Here, the electromechanical linewidth narrowing induced by the driving field cancels the internal losses of the mechanical system, leading to a regime of zero damping (i.e. parametric instability) which is not discussed here.

\begin{figure}[tb]
\center{\includegraphics[width=0.8\columnwidth]{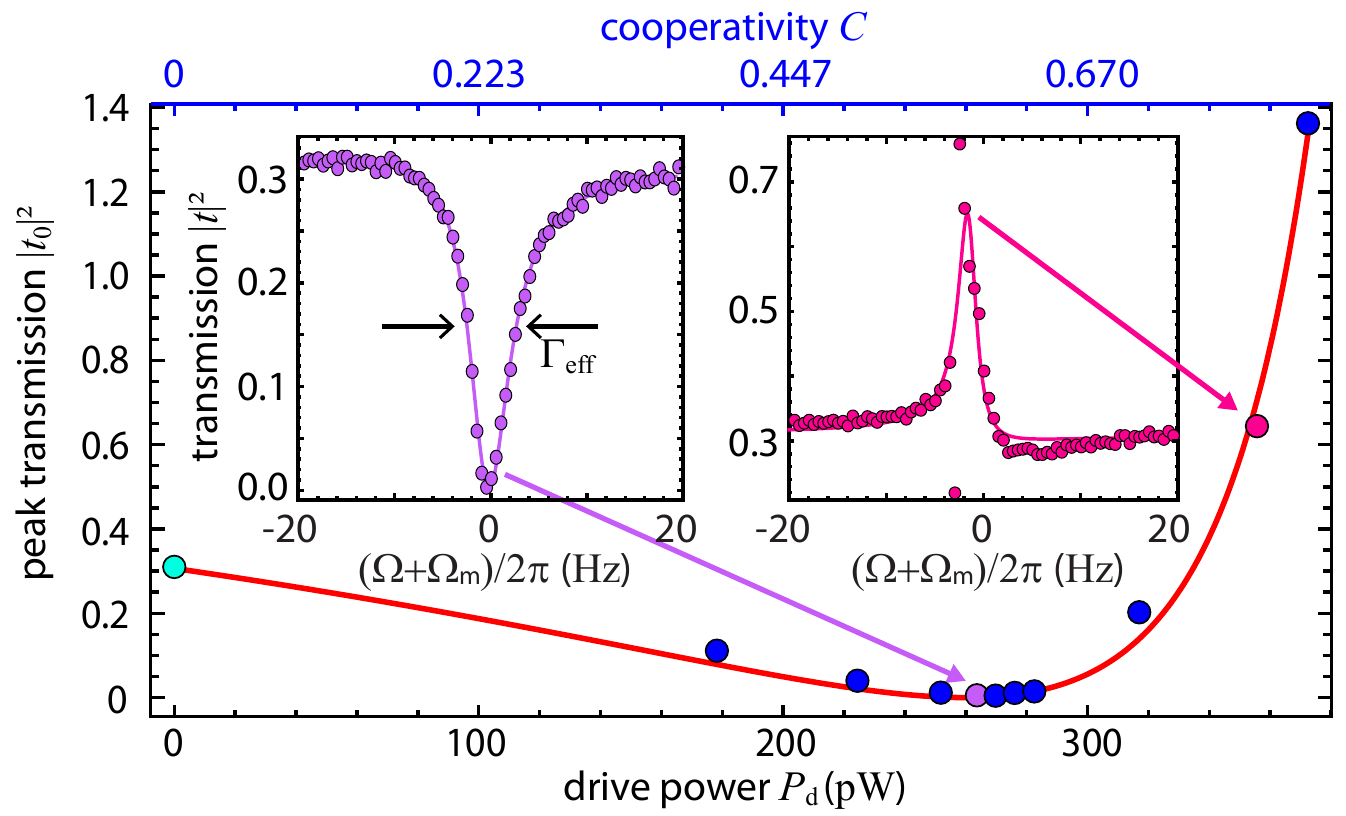}}
 \caption{
  Measured peak probe power transmission $|t_0|^2$ at optimum drive tone detuning $\Delta = \Omega_{\rm m}$ as a function of the drive power $P_{\rm d}$. The green data point shows the power transmission of the bare cavity without a drive tone. The red line is a fit of the data to eq.(\ref{equ:S21_max}). In the left (right) inset, the probe power transmission $|t|^2$ is plotted versus the probe tone detuning $\Omega + \Omega_{\rm m}$ for a constant drive power of $264$\,pW, ($356$\,pW) showing the typical EMIA dip (peak) at $\Omega = -\Omega_{\rm m}$. The symbols represent the measured data, the solid line is obtained by fitting the data by a damped harmonic oscillator spectrum. The fitting process yields $\left|t_{\rm{0}}\right|^2$ at $\Omega = - \Omega_{\rm m}$ for a particular drive power. These values represent the data points in the $|t_0|^2$ versus $P_{\rm d}$ dependence. }
\label{Hocke_EMIA_Figure3}
\end{figure}

An important feature of EMIA is the fact that without additional adjustments the variation of the drive power by a factor of about two leads to a variation of the probe power transmission of almost three orders of magnitude.
Qualitatively, the functional dependence of $\left|t_{\rm{0}}\right|^2$ on the drive power can be understood in terms of the interference of the probe tone and the part of the drive tone, which is down-converted and coherently scattered into the cavity due to phonon generation. If the drive power is low enough, the number of down-converted drive photons is smaller than the number of probe photons sent to the cavity. Therefore, the stimulated absorption of probe photons by the cavity caused by constructive interference with the scattered drive photons is only partial. Note that the physics is completely analogous to EMIT, where destructive interference of the probe and drive tone results in stimulated emission of the cavity, causing an enhanced probe power transmission. With increasing drive power, the absorption dip deepens and minimum transmission is obtained if the number of down-converted drive photons becomes equal to the number of probe photons. With further increasing drive power the number of down-converted drive photons exceeds the number of probe photons, resulting in an increase of the probe power transmission above the level of the bare cavity. That is, the absorption dip in the transmission spectrum changes sign and turns into a transmission peak. We point to the fact that the appearance of a peak in the transmission signal should not be mistaken for an induced transparency effect in the sense of EMIT. At even larger drive power, the power transmission exceeds unity due to parametric amplification of the probe tone by the electromechanical system, which can be considered as a heavily pumped parametric amplifier.

\begin{figure}[tb]
\center{\includegraphics[width=1.0\columnwidth]{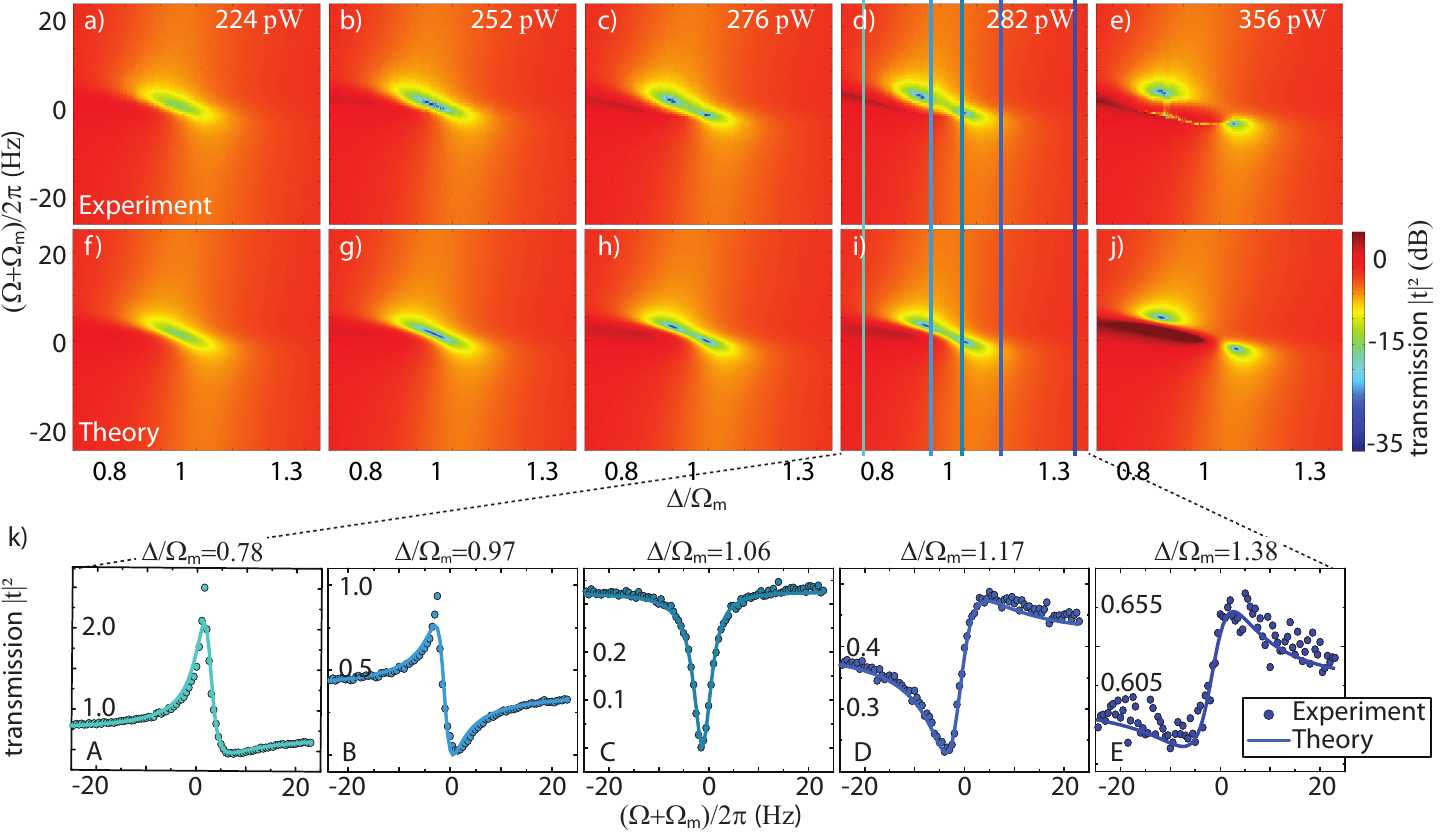}}
 \caption{
  (a)-(e) Color-coded representation of the probe power transmission $|t|^2$ as a function of the normalized drive tone detuning $\Delta/\Omega_{\rm m}$ and the probe tone detuning $\Omega+\Omega_{m}$ away from the Stokes line. The constant drive field power, which is increasing from (a) to (e), is indicated in the upper right corner of each experimental dataset. (f)-(j) Model fits to the data. (k) Vertical line scans along the probe tone frequency axis for various fixed values of the drive detuning $\Delta/\Omega_{\rm m}$ at a constant drive power of 282\,pW. The symbols represent the data, the solid lines are model fits to the data according to eq.(\ref{equ:S21_approx}).}
 \label{Hocke_EMIA_Figure4}
\end{figure}

A quantitative analysis of the measured probe power transmission shown in figure~\ref{Hocke_EMIA_Figure3} can be made by eq.(\ref{equ:S21_approx}), which can be rewritten for optimal drive tone detuning $\Delta = \Omega_{\rm m}$ and $\Omega=-\Omega_{\rm{m}}$ as
\begin{equation}
\left| t_{\rm{0}} \right|^2 = \left| \frac{1-(\kappa_{\rm{ex}}/\kappa) -C}{1-C} \right|^2 \; .
\label{equ:S21_max}
\end{equation}
The red solid line in figure~\ref{Hocke_EMIA_Figure3} shows a fit of this expression to the data. Evidently, the measured data can be well described in a quantitative way. Replacing $\bar{n}_{\rm c}$ in the fitting parameter $C = 4g_{0}^2\bar{n}_{\rm c}/\kappa\Gamma_{\rm{m}}$ by eq.(\ref{equ:photonnum}), the vacuum electromechanical coupling $g_{0}$ can be derived, yielding $g_{0}/2\pi = 1.22$\,Hz. This corroborates the value of $g_{0}/2\pi = 1.26$\,Hz determined by independent frequency noise calibration \cite{Gorodetsky2010a, Zhou2012}. The data analysis in terms of eq.(\ref{equ:S21_max}) is valuable in two respects. On the one hand, it allows for a quantitative explanation of the drive power dependence of the probe transmission. On the other hand, fitting the data by eq.(\ref{equ:S21_max}) yields the cooperativity $C$, which ranges from $0.38$ to $0.80$ for the studied drive power range. These values show that the studied system is on the edge to the strong coupling regime. However, self-oscillation sets in before reaching this regime.

Up to now, we have only discussed the case of optimal drive tone detuning $\Delta = \Omega_{\rm m}$. In the following this analysis is extended to the case of arbitrary detunings $\Delta \neq \Omega_{\rm m}$. Figure~\ref{Hocke_EMIA_Figure4}(a) shows a two-dimensional color-coded representation of the probe power transmission $|t|^2$ around the Stokes line as a function of the drive detuning $\Delta = \omega_{\rm p} - \omega_{\rm c}$, while keeping the probe tone window centered around $\omega_{\rm{p}} = \omega_{\rm{d}}-\Omega_{\rm m}$ (equivalent to $\Omega = -\Omega_{\rm m}$). That is, the frequency window of the probe tone has a constant detuning from the drive tone, thereby probing the same frequency window around the Stokes line [cf. red-shaded area in Fig.~\ref{Hocke_EMIA_Figure1}(a)]. The variation of the drive tone detuning $\Delta$ means that the Stokes line and the related probe frequency window are shifted with respect to the cavity resonance [cf. grey-shaded area in figure~\ref{Hocke_EMIA_Figure1}(a)]. The probe and drive powers are adjusted to $2.4$\,fW and $224$\,pW, respectively. By the two-dimensional spectroscopy shown in figure~\ref{Hocke_EMIA_Figure4}(a), we obtain complete information on the probe power transmission. We first discuss the variation of $|t|^2$ along the drive tone detuning $\Delta/\Omega_{\rm m}$ (horizontal axis) at $\Omega + \Omega_{\rm m}=0$. At a relative drive tone detuning $\Delta/\Omega_{\rm m}=1$ (optimal detuning), we find the absorption feature already discussed above. When the drive detuning is not optimal, $\Delta/\Omega_{\rm m} \neq 1$, we observe that the absorption feature decreases and becomes vanishingly small when moving away from optimal drive detuning. This is obvious, since we are shifting the Stokes line out of the resonance window of the microwave cavity, which acts as a narrow-band filter of width $\kappa$ [cf. figure~\ref{Hocke_EMIA_Figure1}(a)]. Outside this window the absorption feature is strongly suppressed. We note that $\kappa/\Omega_{\rm m} \simeq 0.5$, in good agreement with the width of the absorption feature along the $\Delta/\Omega_{\rm m}$ axis. The panels in figure~\ref{Hocke_EMIA_Figure4}(a) to (e) show the transmission feature for increasing drive power. The absorption at optimal detuning increases until it reaches the largest value at a drive power of $252$\,pW [cf. figure~\ref{Hocke_EMIA_Figure4}(c)]. For even higher drive power, the transmission at optimal detuning increases again, as already discussed in the context of figure~\ref{Hocke_EMIA_Figure3}.

We next discuss the structure of the absorption feature along the probe tone detuning $\Omega + \Omega_{\rm m}$ (vertical axis). For optimal drive tone detuning, $\Delta/\Omega_{\rm m}=1$, the same probe power transmission curve as already shown in the inset of figure~\ref{Hocke_EMIA_Figure3} is obtained. However, when moving away from optimal drive tone detuning, $\Delta/\Omega_{\rm m} \neq 1$, we observe that the absorption feature changes its shape. This becomes most obvious in figure~\ref{Hocke_EMIA_Figure4}(k), where we have plotted the probe power transmission as a function of the probe tone detuning (vertical line scans) for several fixed values of the drive detuning $\Delta/\Omega_{\rm m}$ marked by the vertical lines in figures~\ref{Hocke_EMIA_Figure4}(d) and (i). The observed behavior can be attributed to the CPW resonator, whose envelope can be considered as a dispersive background.  For $\Delta/\Omega_{\rm m} = 1.06$, the Stokes line and probe frequency window is positioned close to the center of the cavity resonance. In this case the phase shift between the probe and cavity field is about $\pi/2$, whereas it changes towards $0$ and $\pi$ upon moving away from optimal drive detuning $\Delta/\Omega_{\rm m} = 1$, i.e. when shifting the Stokes line and the probe window out of the center of the cavity resonance. This results in transmission curves resembling the imaginary and real part of the susceptibility of a damped harmonic oscillator for the probe frequency window positioned close to and left/right from the resonance, respectively. The data shown in figure~\ref{Hocke_EMIA_Figure4}(k) can be well approximated by a Lorentzian in a dispersive environment.

\begin{figure}[tb]
\center{\includegraphics[width=0.6\columnwidth]{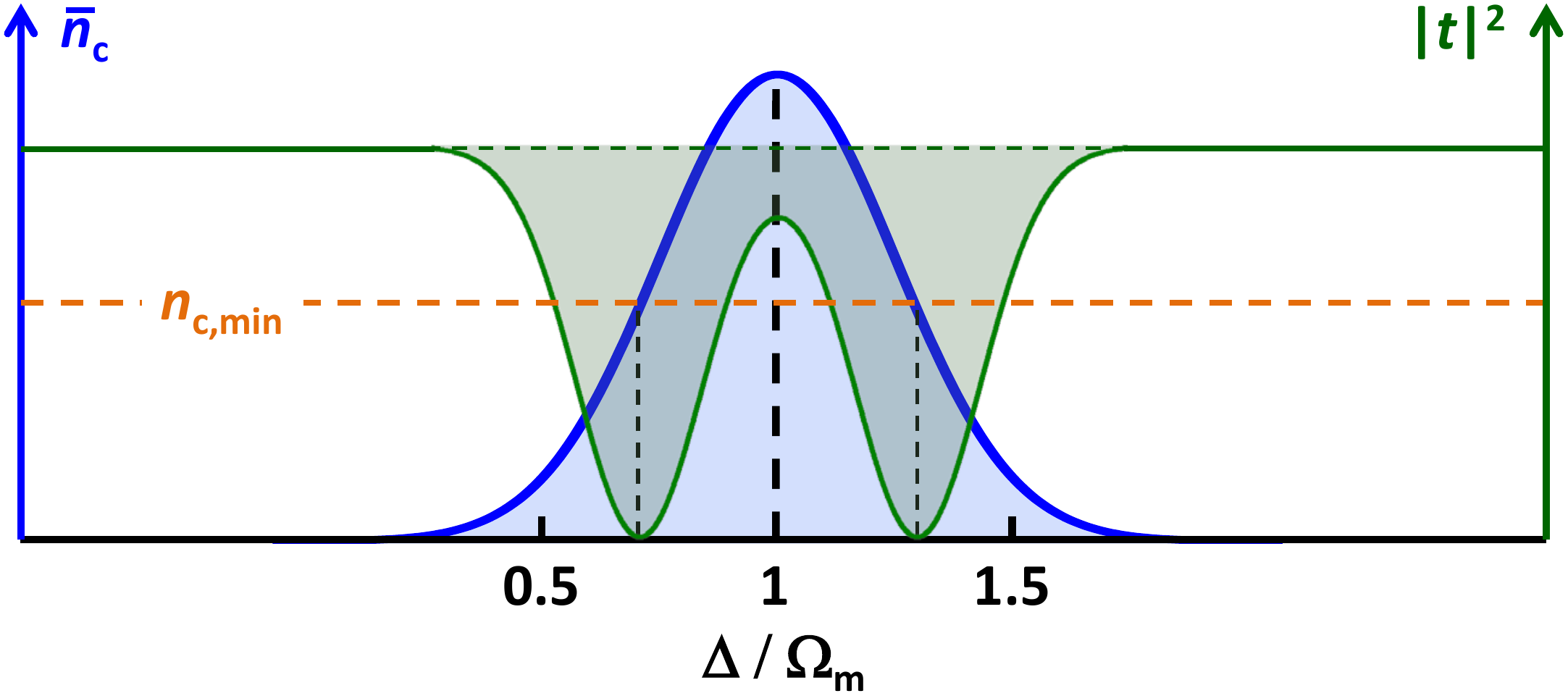}}
 \caption{
  Schematic diagram showing the variation of the average cavity photon number $\bar{n}_{\rm c}$ (blue) and the resulting probe power transmission $|t|^2$ (olive) as a function of the drive tone detuning $\Delta/\Omega_{\rm m}$ at constant applied drive tone power. The photon number $\bar{n}_{\rm c}$ is assumed to vary as the cavity transmission. Minima in the probe power transmission are obtained for $n_{\rm c}(\Delta/\Omega_{\rm m})=n_{\rm c,min}$, resulting in the characteristic double dip structure observed in the experiment. }
 \label{Hocke_EMIA_Figure5}
\end{figure}

We also observe that the single transmission dip around $\Delta/\Omega_{\rm m} = 1$ and $\Omega + \Omega_{\rm m}=0$ splits into two along the drive frequency detuning on increasing the drive power. This behavior is not a signature of the normal mode splitting \cite{Dobrindt2008, Metzger2008}, setting in when the coupling strength $g = g_0 \sqrt{\bar{n}_{\rm c}}$ becomes comparable to the cavity loss rate $\kappa$. However, for our system, $g \ll \kappa$ even for the largest driving powers. The double dip feature results from the effect that the number of down-converted drive photons varies with the drive tone detuning $\Delta/\Omega_{\rm m}$ due to the filtering function of the cavity as sketched in figure~\ref{Hocke_EMIA_Figure5}. Since the minimum transmission is achieved for a specific cavity photon number $n_{\rm c,min}$ and increases both for smaller and larger values, two dips in the probe power transmission are obtained at those drive detunings where $n_{\rm c}(\Delta/\Omega_{\rm m})=n_{\rm c,min}$. Finally, for the highest drive power of $356$\,pW [cf. figure~\ref{Hocke_EMIA_Figure4}(e)] we find additional sharp features forming a line of very narrow dips. They can be attributed to the transition to the nonlinear Duffing regime \cite{Zhou2012}.

To analyze the absorption more quantitatively, one has to compare the experimental results with eq.(\ref{equ:S21_approx}). Figures~\ref{Hocke_EMIA_Figure4}(f) to (j) show the probe power transmission $|t|^2$ obtained according to this equation. Evidently, the results of the model calculation agree well with the experimental data, apart from the onset of nonlinear features for high driving power. Additionally, figure~\ref{Hocke_EMIA_Figure4}(k) displays vertical line scans along the probe tone frequency axis for a drive power of $141$\,pW. Fitting the data yields  $\Omega_{\rm{m,fit}}/2\pi = 1.45$\,MHz and $g_{\rm{0,fit}}/2\pi = 1.29$\,Hz, corroborating the values given above.

\section{Conclusion}

In conclusion, we have performed a detailed two-tone spectroscopy analysis of electromechanically induced absorption (EMIA) in a hybrid system consisting of a superconducting microwave resonator coupled to a nanomechanical beam as function of the drive power. In two-dimensional spectroscopy experiments the probe power transmission has been measured both as a function of the drive and probe tone detuning for a wide range of drive tone powers. We find good quantitative agreement between the measured transmission spectra and model calculations based on the Hamiltonian formulation of a generic electromechanical system. For optimal drive tone detuning we show that the absorption of microwave signals at cavity resonance can be adjusted by more than $25$\,dB on varying the power of the drive tone by a factor of two. A minimum normalized power transmission of $0.0046$ has been demonstrated in a very narrow absorption window of $\Delta\omega/2\pi=5$\,Hz at $\omega=6$\,GHz, resulting from line narrowing in the dressed mechanical system. Even narrower band pass filters have be achieved at larger drive powers, however with increased probe power transmission. Our results clearly demonstrate that the studied electromechanical system can be applied to filter out extreme narrow frequency bands ($\sim$\,Hz) of the much wider frequency band ($\sim$\,MHz) defined by the linewidth of the microwave cavity. The amount of absorption as well as the filtering frequency is tunable around the cavity resonance over about $1$\,MHz by adjusting the power and frequency of the drive field.

Another possible application of the nanomechanical system is the nonlinear manipulation of light fields down to the quantum level by introducing a group advance or delay to weak microwave pulses centered around the cavity resonance \cite{Zhou2012}. This capability results, similar to the case of EMIT, from the rapid phase dispersion originating from EMIA. In contrast to EMIT, for EMIA the change from  absorptive to transmissive behavior allows both for advancing and delaying microwave signals.

At high drive power we observed parametric amplification of the weak probe tone, that is, microwave amplification with a nanomechanical resonator. However, due to the low mechanical loss rate of the nanomechanical beam ($\Gamma_{\rm m}=2\pi\times11$\,Hz) the drive power range for parametric amplification is very narrow, since the beam rapidly starts to perform self-oscillations and eventually shows phonon lasing. Experiments exploring this regime are in progress.

\ack
The authors gratefully acknowledge financial support from the German Excellence Initiative via the ``Nanosystems Initiative Munich`'' (NIM). T.J.K acknowledges support by the ERC grant SIMP and X.Z. by the NCCR of Quantum Engineering. Samples were grown and fabricated at the Center of MicroNanotechnology (CMi) at EPFL.

\section*{References}

\bibliographystyle{unsrt}

\end{document}